\documentclass[a4paper]{PoS}
\usepackage{graphicx,wrapfig,lipsum}
\def\VYP#1#2#3{{\bf #1}, #3 (#2)}  

\title{On the feasibility of RADAR detection of high-energy cosmic neutrinos}

\ShortTitle{On the feasibility of RADAR detection of high-energy cosmic neutrinos}

\author{\speaker{K.D. de Vries}\\
        Vrije Universiteit Brussel, Dienst ELEM, B-1050 Brussels, Belgium\\
        E-mail: \email{krijndevries@gmail.com}}

\author{K. Hanson\\
        UW Wisconsin, Madison, Wisconsin\\
        Universit\'e Libre de Bruxelles, Department of Physics, B-1050 Brussels, Belgium
        }

\author{T. Meures\\
        UW Wisconsin, Madison, Wisconsin\\
        Universit\'e Libre de Bruxelles, Department of Physics, B-1050 Brussels, Belgium
        }

\author{A. \`O Murchadha\\
        Universit\'e Libre de Bruxelles, Department of Physics, B-1050 Brussels, Belgium}

\abstract{We discuss the radar detection technique as a probe for high-energy cosmic neutrino induced particle cascades in a dense medium like ice. With the recent detection of high-energy cosmic neutrinos by the IceCube neutrino observatory the window to neutrino astronomy has been opened. We discuss a new technique to detect cosmic neutrinos at even higher energies than those covered by IceCube, but with an energy threshold below the currently operating Askaryan radio detectors. A calculation for the radar return power, as well as first experimental results will be presented.}

\FullConference{The 34th International Cosmic Ray Conference,\\
		30 July- 6 August, 2015\\
		The Hague, The Netherlands}

\begin{document}

\section{Introduction}
With the recent discovery of cosmic neutrinos with energies up to several PeV by the IceCube collaboration~\cite{I3_2013sc}, the window to neutrino astronomy has been opened. Nevertheless, above several PeV an even larger effective volume than the cubic kilometer filled by IceCube is needed. Due to its long attenuation length, of the order of a kilometer in ice, the radio signal is an excellent probe for the detection of high-energy cosmic neutrinos at even higher energies. 

The direct radio detection technique was predicted by Askaryan in the 1960's and is based on the coherent emission from a net negative charge excess which arises when the neutrino induced particle cascade develops~\cite{Ask62}. The currently operating Askaryan radio detectors at the South Pole~\cite{ARA,ARIANNA,ANITA} start to become sensitive at energies of several EeV where the GZK-flux is expected. We present a method to lower this energy threshold to several PeV by means of the radar detection technique. The radar detection of cosmic-ray induced air showers was already proposed in the previous century~\cite{Bla40}, and revived in the beginning of this century~\cite{Gor01}, we discuss this technique to probe particle cascades in a dense medium like ice.

In~\cite{dVries15}, an energy threshold of a few PeV was determined for the radar detection of the over-dense ionization plasma induced when the cascade propagates through ice. A conservative estimate was given for the detection distance which was of the order of a few hundreds of meters up to several kilometers depending on the considered geometry and the properties of the plasma. We briefly recall the derivation of this threshold and discuss the remaining uncertainties. These uncertainties are mainly due to the different properties of the ionization plasma, such as its lifetime, and the electron collision frequency. To obtain a better handle on the ionization plasma an experiment was performed testing the radar scattering off of a plasma induced by shooting a beam of electrons in a block of ice at the Electron Light Source at the Telescope Array site~\cite{Shi08}. In these proceedings we present the first results of this experiment. 

\section{Over-dense radar scattering}
We discuss the radar scattering off of an over-dense ionization plasma which is induced when a high-energy particle cascade propagates through a dense medium. Over-dense scattering is obtained through the condition that the observation frequency has to be below the plasma frequency,
\begin{equation}
\nu_{obs} < \nu_{p}.
\end{equation}
In this situation, an electromagnetic wave will scatter off of the full plasma volume. At observation frequencies above the plasma frequency, the wave will not scatter off of the collective volume, but rather off of the individual free charges in the plasma. This process occurs through Thomson scattering greatly reducing the radar cross-section in case of under-dense scattering. 

The plasma frequency is a property of the plasma directly linked to the free charge density $n_e\;\mathrm{[cm^{-3}]}$,
\begin{equation}
\nu_{p}=8980\sqrt{\frac{m_e}{m_p}n_e},
\end{equation}
where $m_p$ denotes the (effective) mass of the plasma constituent, and $m_e$ is the electron mass. In~\cite{dVries15}, we give an estimate for the ionization electron density induced by a high-energy particle cascade in ice. It was shown that the electron density can be linked directly to the energy of the primary neutrino,
\begin{equation}
n_e\approx 3.5\cdot 10^3 E_p\;\mathrm{[GeV]\;cm^{-3}}.
\end{equation}
The plasma frequency can now be rewritten as function of the energy op the primary neutrino,
\begin{equation}
\nu_{p}\approx0.5\sqrt{\frac{m_e}{m_p}E_p\;\mathrm{[GeV]}}\;\mathrm{MHz}.
\end{equation}
The second condition for scattering off of the over-dense plasma is determined by either the lifetime, $\tau_p$, or the dimensions $l_c$ of the plasma. The observation frequency is bound by the condition,
\begin{equation}
 \nu_{obs} >\left\{
 \begin{array}{l l}
 1/\tau_p\;\;\;\;\;\;\;\;\;(c_{med}\tau_p < l_c)\\
 c_{med}/l_c\;\;\;\;(c_{med}\tau_p > l_c)
 \end{array}
 \right.\;,
\end{equation}
where $c_{med}$ denotes the speed of light in the medium. In~\cite{dVries15}, the dimensions of the induced plasma are modeled and found to be of the order of several meters. The larger uncertainty lies in the lifetime and constituent of the plasma. For this we consider measurements performed in the 1980s~\cite{Ver78,Kun80}, where a block of ice was irradiated with an electron beam as well as X-rays. In these measurements, two different plasma constituents were found. The first plasma was attributed to the ionization electrons, and lifetimes of the order of 100~ps at relatively high temperatures to tens of nanoseconds at $-60^\circ$ degrees Celcius were obtained. Next to the ionization electrons, a plasma with the properties equal to a free proton plasma was obtained with much longer lifetimes ranging from 10~ns to 1~$\mathrm{\mu s}$.

Since ice temperatures at natural ice-sheets like those at the South Pole are found to be of the order of $-50^\circ$, for the electron plasma we take a still conservative value of $\tau_e=1$~ns, leading to a limit on the observation frequency of $\nu_{obs}^{e}>1$~GHz. For the proton plasma, the observation frequency will be limited by the dimensions of the plasma. Taking $l_c=5$~m, the observation frequency is limited by $\nu_{obs}^{p}>36$~MHz.  

Combining Eqs.~2.1), 2.4), and 2.5) we can now put an energy threshold for the over-dense scattering off of the induced ionization plasma. For the ionization-electron plasma, taking a plasma frequency equal to the minimum observer frequency of 1~GHz, we obtain an energy threshold of $E^{e}\gtrsim 4$~PeV. For the second constituent which has the properties of a free proton plasma, this becomes $E^{p}\gtrsim 20$~PeV. 

\section{The radar return power}
\begin{figure}[ht]
\centering
\begin{minipage}{.45\textwidth}
  \centering
  \includegraphics[height=.75\textwidth]{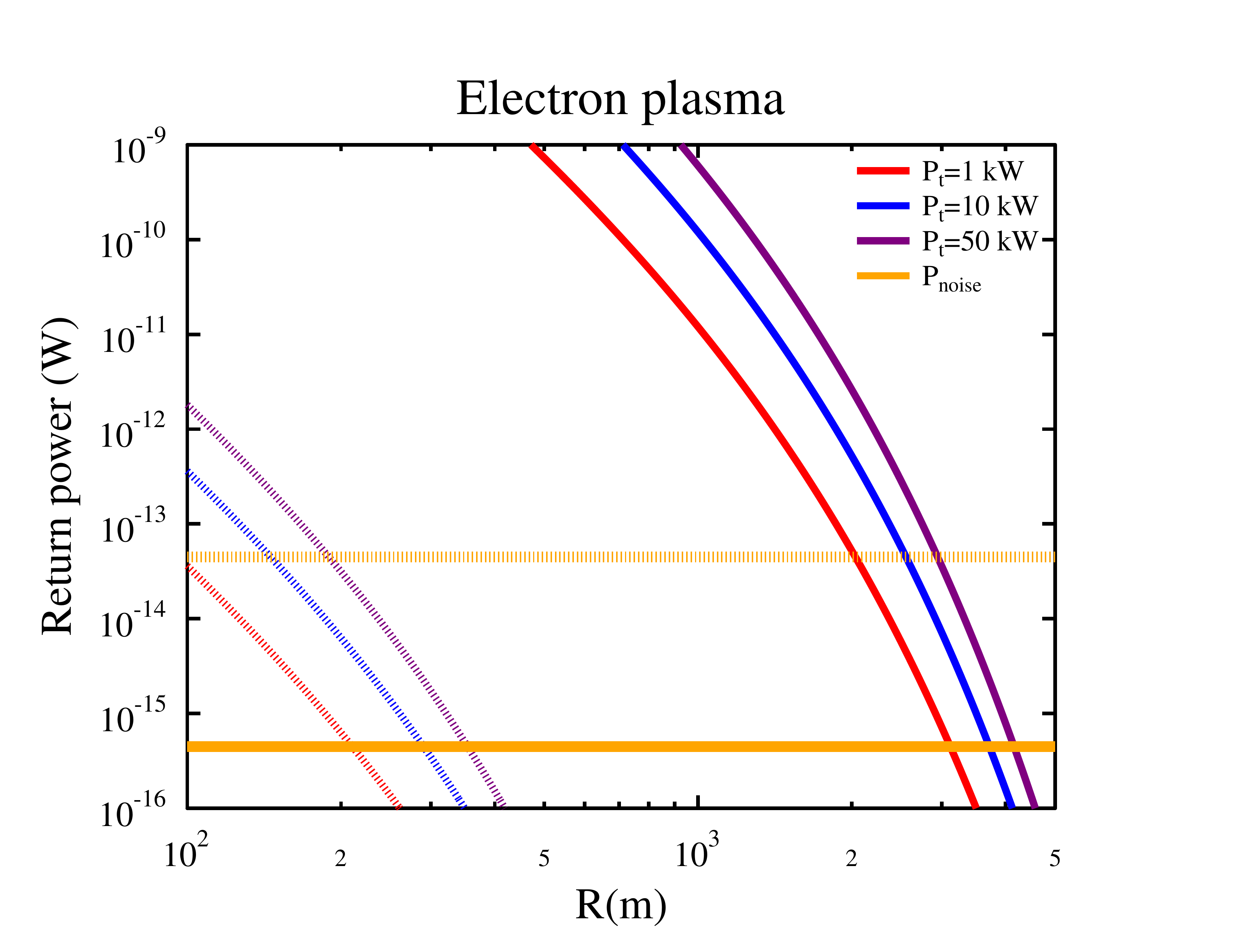}
  \label{fig:test1}
\end{minipage}%
\begin{minipage}{.45\textwidth}
  \centering
  \includegraphics[height=.75\textwidth]{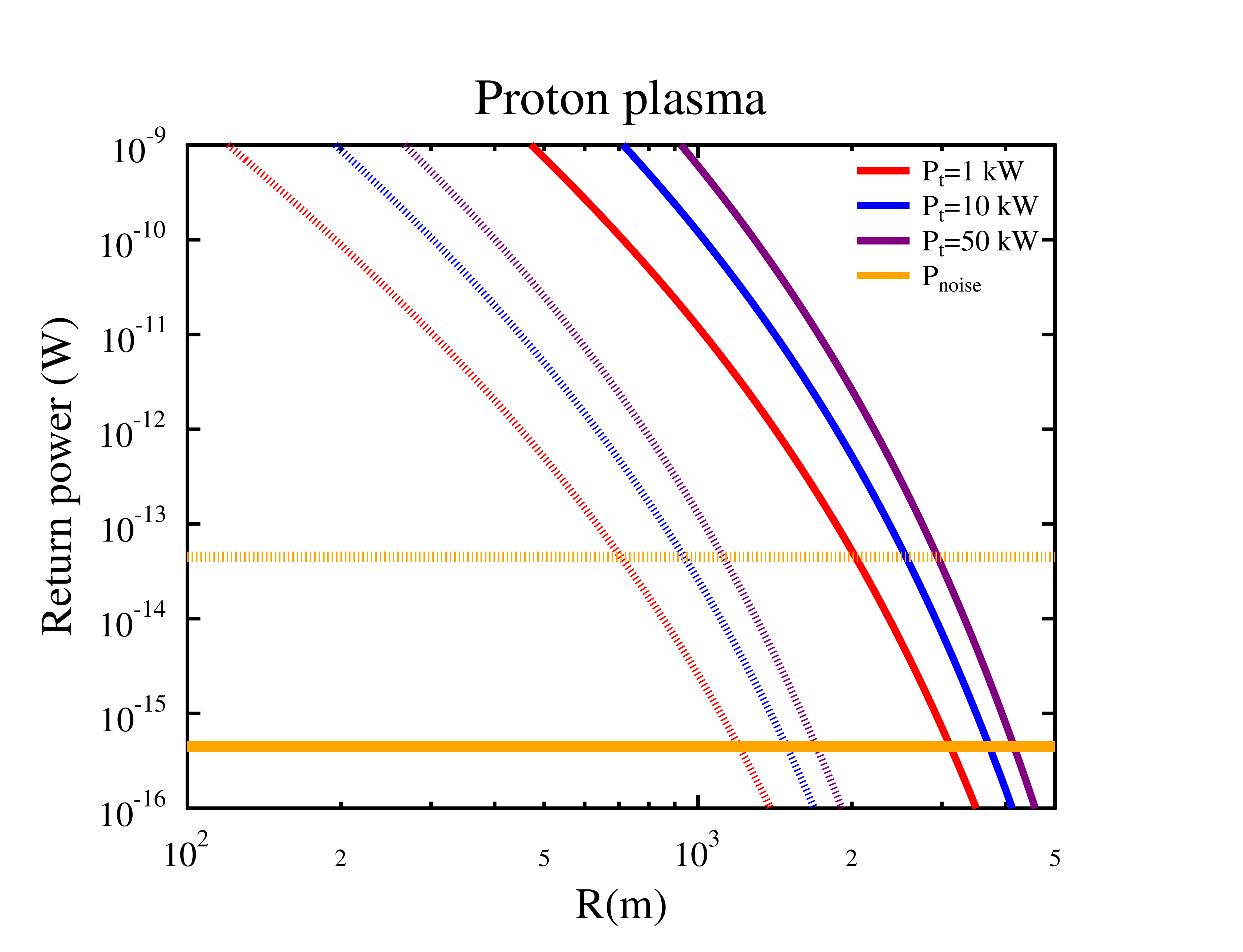}
\end{minipage}
\label{fig1}
\caption{The radar return power as a function of the source distance for the electron plasma (left) and the proton plasma (right). The return power is given for two different situations, the most conservative approximation shown by the dashed lines and a more progressive estimate given by the full lines.}
\end{figure}

In~\cite{dVries15} we considered the radar return power for both the electron plasma as well as the proton plasma. In calculating the radar return power several parameters were fixed rather conservatively. In this section we discuss these parameters and consider two different situations at both the conservative end of the parameter space, as well as the progressive side of the parameter space. Following~\cite{Gor01}, the radar return power for a bi-static radar configuration is given by,
\begin{equation}
P_r=P_t\eta\frac{\sigma_{eff}}{\pi R^{2}}\frac{A_{eff}}{4\pi R^2}e^{-4R/L_\alpha}.
\end{equation}
Here we consider a symmetric configuration such that the distance $R$ from the emitter to the cascade is equal to the distance from the cascade to the receiving antenna. The different parameters in this equation are given in Table~\ref{Tab1}.
\begin{table}
\centering
\begin{tabular}{ll}
$\eta$ & 0.1 \\
$\sigma^{e}_{eff}(\nu=0.05-1~\mathrm{GHz},\theta=0^\circ-60^\circ,\phi=0^\circ-60^\circ)$  & $1.6\cdot10^{-4}-5.5\;\mathrm{m^2}$ \\
$\sigma^{p}_{eff}(\nu=50~\mathrm{MHz},\theta=0^\circ-60^\circ,\phi=0^\circ-60^\circ)$  & $1.2\cdot10^{-2}-5.5\;\mathrm{m^2}$\\
$L_\alpha^e(\nu=0.05-1~\mathrm{GHz})$  & 313-1500~m\\
$L_\alpha^p(\nu=50~\mathrm{MHz})$  & 800-1500~m\\
$T_{sys}$  & 325~K\\
$\Delta \nu$ & 0.1-10~MHz\\
\hline
\end{tabular}
\caption{Parameters used to determine the radar return power and the noise power.}
\label{Tab1}
\end{table}
An overall efficiency factor of $\eta=0.1$ is used to take into account for possible effects that have been ignored so-far. For the radar cross-section, we base ourselves on a thin wire approximation used in~\cite{Gor01,Cri65}. Logically, the radar cross-section is a function of the dimensions of the plasma tube and the detection frequency. Where for the proton plasma, the dimensions are dominated by the length of the cascade, for the electron plasma this is only the situation for the longest lifetimes of several tens of nanoseconds measured at temperatures below $-50^\circ$ Celcius. For the longest lifetimes, the detection frequency is chosen to be $\nu=50$~MHz for both the electron as well as the proton plasma. On the conservative side a lifetime of 1~ns is considered for the electron plasma. In this situation, the detection frequency has to be in the GHz range, and the dimensions of the plasma are dominated by its lifetime. Furthermore, the radar cross-section depends strongly on the polarization angle $\theta$ and incident angle $\phi$ of the incoming wave. The attenuation length is determined by measurements performed by the ARA collaboration at the South-Pole~\cite{ARA}. On the progressive side we use the attenuation length measured at the South-Pole within the first 1500~m of the ice-sheet at a frequency of 300~MHz, which was found to be around 1500 meters. On the conservative side, assuming a low detection frequency of 50~MHz, we took the all-ice value of 800~m as measured by the ARA collaboration at a frequency of 300~MHz. For a high detection frequency of 1~GHz, we use an extremely conservative value for the attenuation length of 313~m which is observed at the Ross ice-shelf~\cite{Bar12}. The noise floor accordingly is dominated by thermal noise and given by $P_{noise}=k_B T_{sys} \Delta \nu$, where the system temperature $T_{sys}=325$~K is chosen following measurements at the ARA site, $k_B$ is Boltzmann's constant and the detection band-width is estimated to lie in between 100~kHz and 10~MHz. 

From Fig.~1 it follows that even in the most conservative situation, for the electron plasma which is being probed under a highly non-ideal geometry, the return power lies above the background for source distances up to a few hundreds of meters. The total distance covered by the received signal is twice the source distance since we consider a symmetric situation where the distance from the transmitter to the plasma, $R$, is equal to the distance from the plasma to the receiver. Considering the proton plasma, this distance increases to several kilometers already in the most conservative approximation. For the more ideal situation considering more progressive parameter values the return power lies above the background for source distances of a few kilometers for both the proton plasma as well as the electron plasma. It follows that the radar detection method is a very promising technique for the detection of high-energy cosmic neutrinos, probing the currently existing energy gap between several PeV and a few EeV.

\section{Radar scattering experiment}
\begin{figure}[h]
\centering
\begin{minipage}{.45\textwidth}
  \centering
  \includegraphics[height=\textwidth]{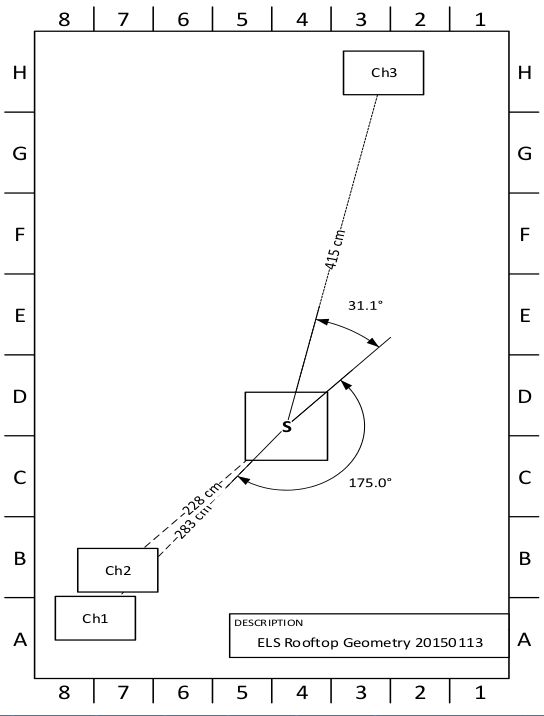}
  \label{fig:test1}
\end{minipage}%
\begin{minipage}{.45\textwidth}
  \centering
  \includegraphics[height=\textwidth]{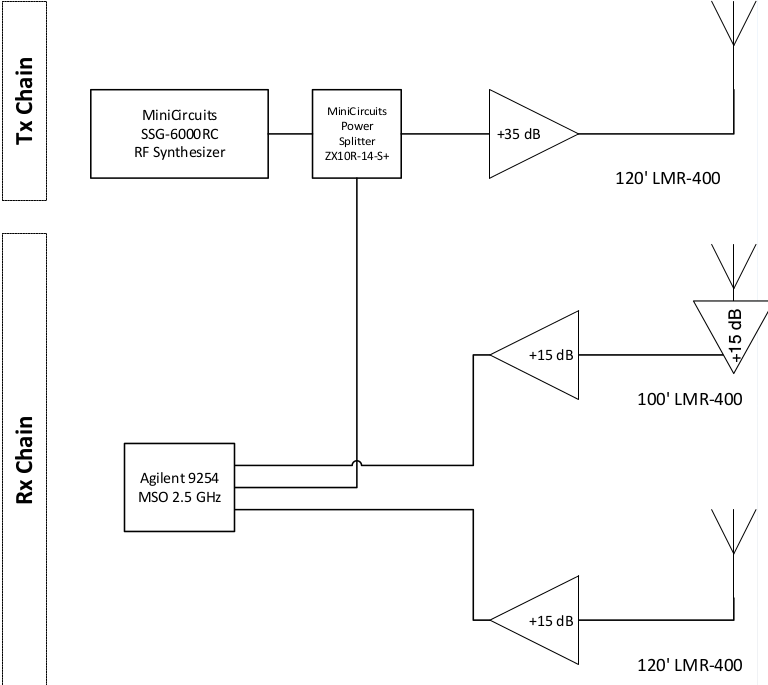}
\end{minipage}
\label{fig1}
\caption{The experimental layout (left), and the signal chain (right) used for the radar scattering experiment at the ELS. The transmitter is positioned at ch2, where the receiver antennas are in the forward direction (ch3), and the backscatter position (ch1).}
\end{figure}

For the calculations performed in the previous section, we considered a fully reflective plasma. Several effects which could lead to a dissipation of energy, like the skin effect or possible energy and coherence loss due to inelastic collisions of the free charges, were absorbed in the efficiency parameter $\eta$. Some of these effects like the free charge collision frequency were modeled in more detail for the radar reflection off of the plasma induced by a cosmic-ray induced air shower~\cite{Sta15}, and shown to be crucial for the radar reflection.

To obtain a better handle on these effects, a first experimental effort was performed to see whether it is possible to bounce a radio wave off of an ionization plasma in ice. In this section we briefly discuss some of the results obtained. A more detailed discussion will be presented in a forthcoming paper.

The experiment was performed at the Electron Light Source (ELS) at the Telescope Array (TA) site~\cite{Shi08}. The ELS is built as a calibration tool for the Fluorescence Detectors of TA, and directs a beam of high-energy electrons upward into the air. For our experiment, a block of ice was positioned on top of this beam. The beam consists of approximately $10^9$, 40~MeV, electrons corresponding to an equivalent energy of 40~PeV. The bunch has a radial extent of approximately 6~cm, and the bunch length is $\sim10$ ns.

In Fig.~2, an schematic layout of the experiment is shown, as well as the signal chain. The data shown in these proceedings are obtained from the reciever antenna in the backscatter direction, denoted by ch2, while transmitting at a frequency of $\nu_{obs}=1550$~MHz. It can be shown that for the given beam configurations a highly over-dense plasma should be induced in the block of ice which is positioned on top of the beam exit point denoted by "S" in Fig~2.

\subsection{Background}
\begin{figure}[h]
\begin{minipage}{.35\textwidth}
  \includegraphics[height=\textwidth]{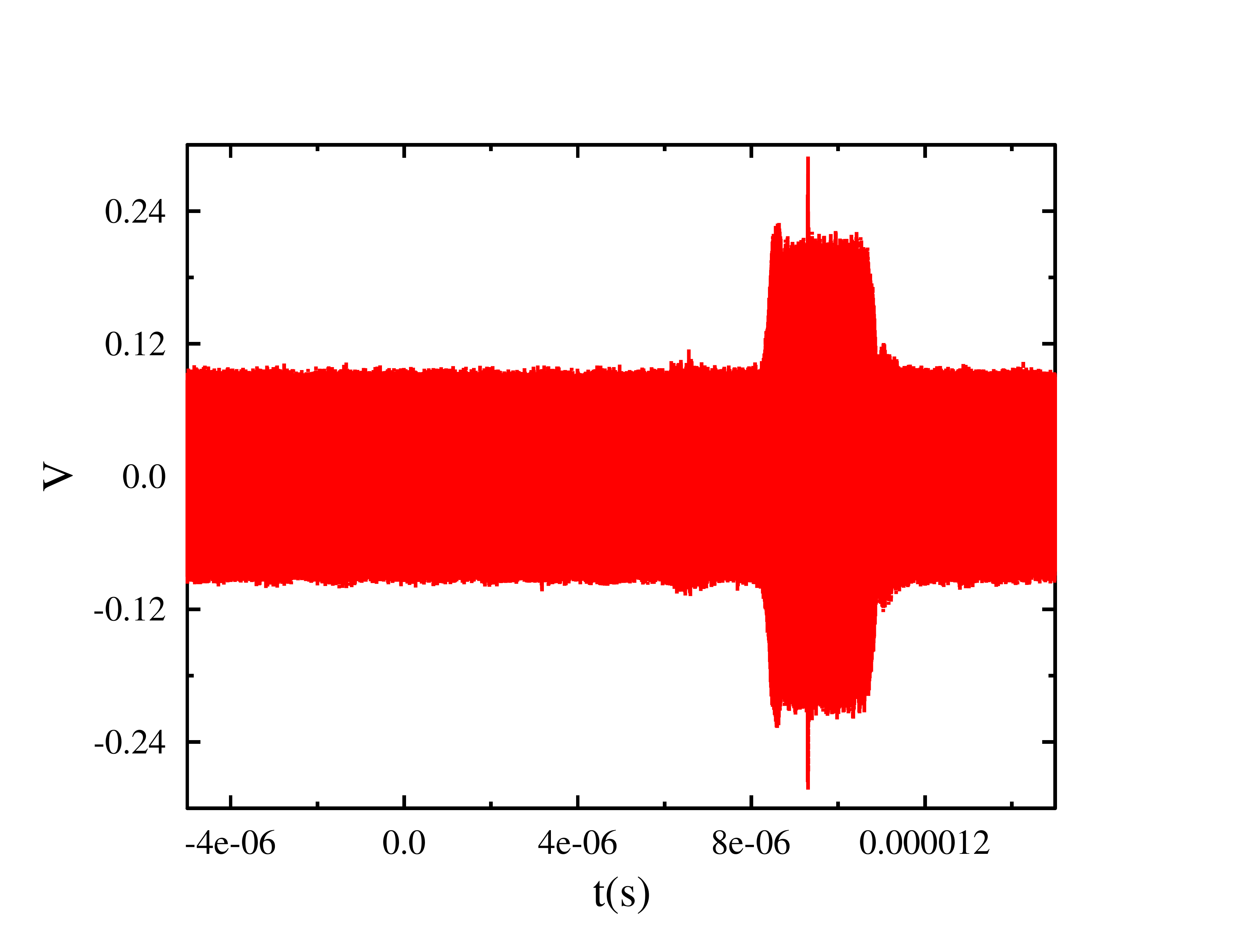}
  \label{fig:test1}
\end{minipage}%
\hspace{1cm}
\begin{minipage}{.35\textwidth}
  \includegraphics[height=\textwidth]{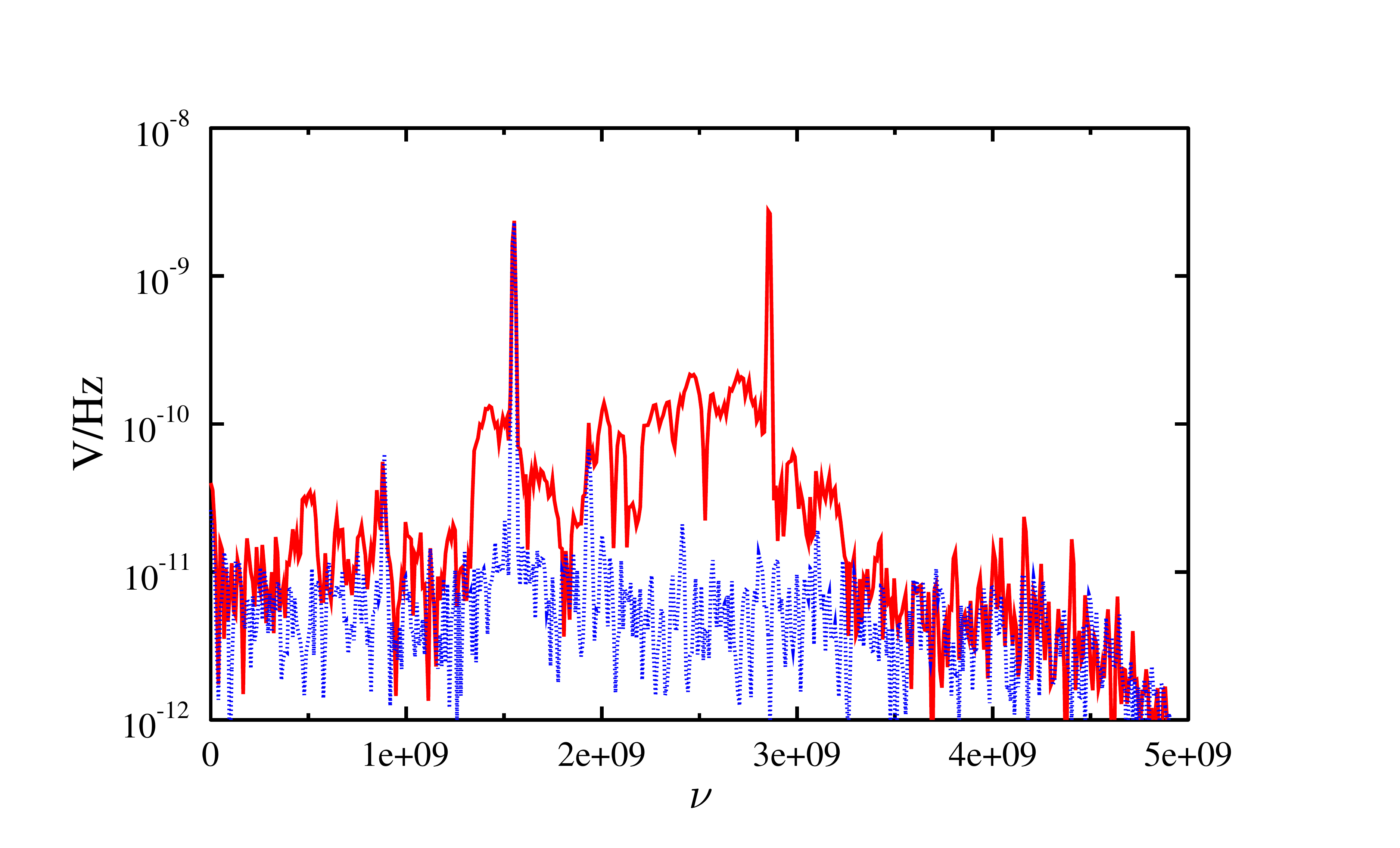}
  \label{fig:test1}
\end{minipage}%

\caption{The raw time trace obtained during the ELS experiment transmitting at $\nu=1550$~MHz (left), and the fourier spectrum (right) in the beam-off region (blue striped line) as well as the beam-on region (full red line).}
\end{figure}


In Fig.~3, we show a typical time-trace obtained during the experiment, along with the fourier spectrum. The fourier spectrum is given in the beam-off configuration (dashed blue line) and the beam-on configuration (full red line). The total trace length is 20~$\mathrm{\mu s}$. Even though our receiver is located outside of the main lobes of the transmitter, direct signal from the transmitter is still relatively strong as can be observed in the first 12~$\mu s$, and from the peak at $1550$~MHz in the fourier spectrum in case there is no beam. After $\sim12\;\mathrm{\mu s}$, the accelerator starts which gives a very strong peak at $2.85$~GHz. Furthermore, in the time-trace there is a strong spike after approximately $\sim14\;\mathrm{\mu s}$ which is a combination of the beam escaping the accelerator giving a so-called sudden-appearance signal, transition radiation from the air-ice boundary, and the Askaryan like signal from the development of the beam in the ice. Also this is the position where we expect our scattered signal, and hence it already follows that a possible scattered signal will be on top of a large background which consists of several components given by, the transmitter signal leaking into the receiver, accelerator noise, and the physical signal due to the beam propagation.

\subsection{Signal}
\begin{figure}[h]
\begin{minipage}{.35\textwidth}
  \includegraphics[height=\textwidth]{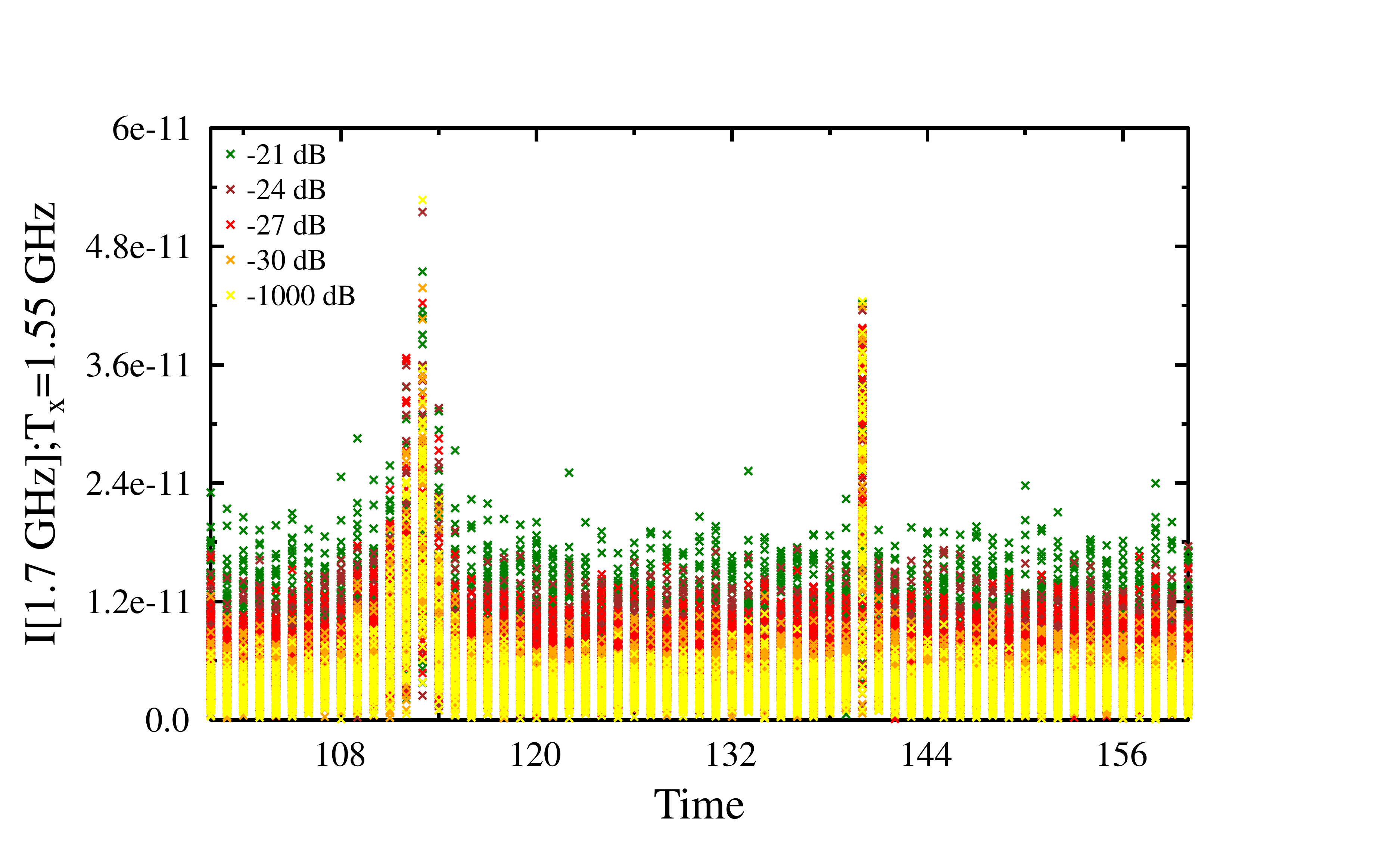}
  \label{fig:test1}
\end{minipage}%
\hspace{2cm}
\begin{minipage}{.35\textwidth}
  \includegraphics[height=\textwidth]{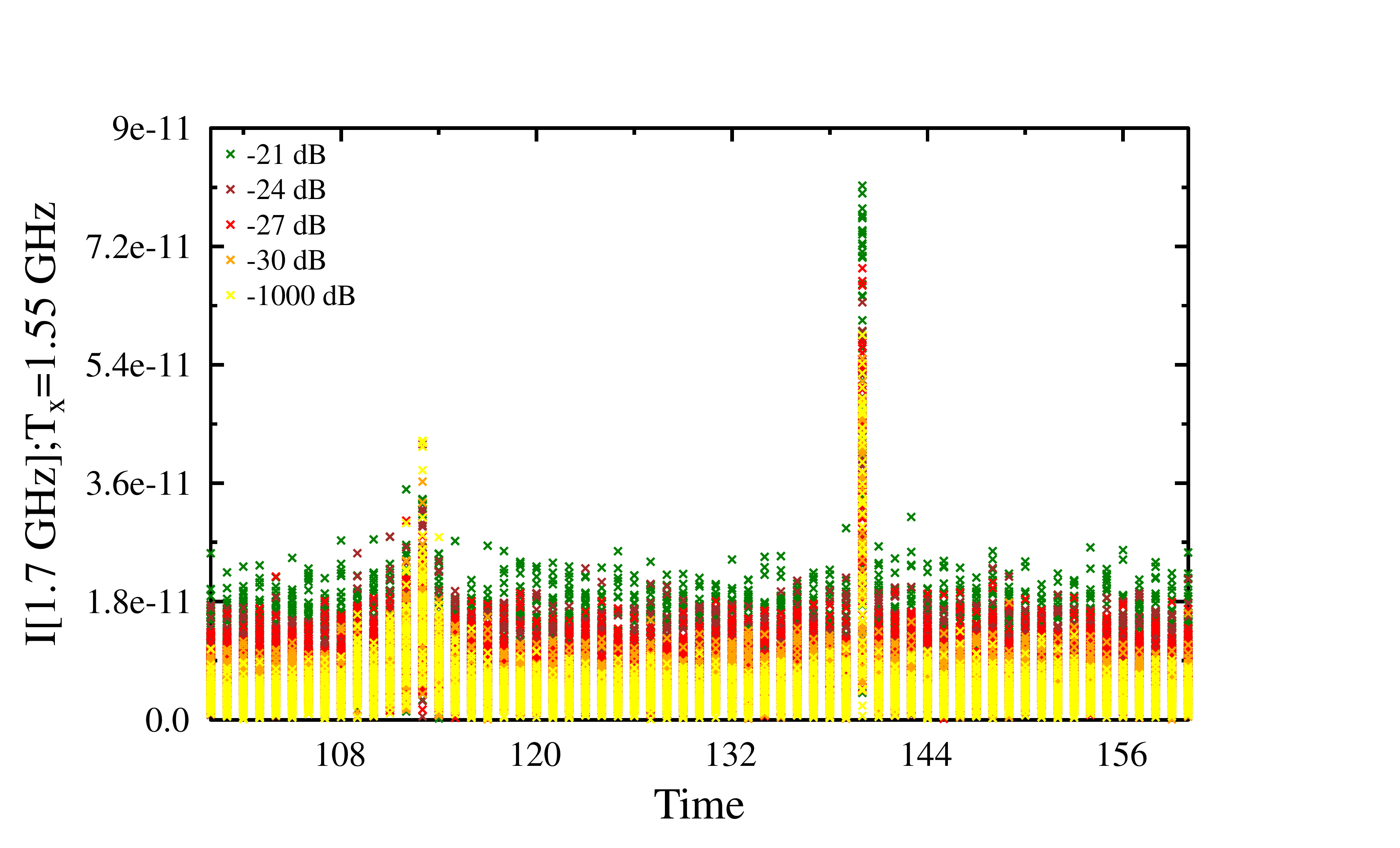}
  \label{fig:test1}
\end{minipage}%
\caption{The intensity at $\nu=1.7$~GHz obtained in different 100~ns long time bins for the no-ice situation where the electron beam was shot into the air (left) and the configuration with the ice-block on top of the beam (right). Each point indicates a the measurement in a 100 ns bin for a single beam shot. In total 50 beam shots for each transmitter intensity are shown. The intensity is indicated by the different colors.}
\end{figure}
One way to avoid noise due to the leakage of the transmitted signal is to observe at a different frequency, slightly away from the transmit frequency. This is shown in Fig.~4, where we show the power of the emission at $\nu=1.7$~GHz. This is done by dividing the raw time-trace into 100~ns time windows. In each of these windows we determine the power at $1.7$~GHz. The field is consequently plotted as function of the time-window number. This is done for 50 consecutive measurements at different values for the transmit power ranging from power off (yellow crosses) to -21~dB (green crosses). 

In the left plot of Fig.~4, we show measurements performed when the block of ice was removed and hence the beam was propagating into the air. From this figure, a noise burst between $10.8\;\mu s$ and $12\;\mu s$ is visible. At $14\;\mu s$, the signal due to the sudden appearance of the beam and propagation of the beam into the air becomes clearly visible. From Eq.~3.1 it follows that for a possible scattered signal, we expect the return power to scale directly with the input power. This is clearly not the situation for the left plot of Fig.~4, where the beam is directed into air. The right plot of Fig.~4, shows the results obtained when the block of ice was positioned on top of the beam. From this figure it clearly follows that suddenly a scaling with the input power is obtained. Even-though this gives a very strong indication for a scattered signal, data obtained at other frequencies is less clear and mostly dominated by the background discussed in the previous section.

\section{Conclusion}
We discussed the feasibility of the radar detection technique as a probe for high-energy neutrino induced particle cascades in ice. We considered the radar scattering off of the static ionization plasma induced when the cascade looses its energy to the medium. Measurements performed in the previous century indicate that two types of plasma have to be considered. The first is a rather short-lived free electron plasma, the second is a relatively long lived plasma with the properties free protons. For the over-dense scattering off of the electron plasma we derived an energy threshold of 4~PeV, while for the proton plasma an energy threshold of 20~PeV was derived. 

Next to the energy threshold we also determined the radar return power giving a conservative as well as a more progressive estimate. It follows that the scattered signal should be visible starting from a few hundreds of meters in the most conservative approach up to several kilometers in the more progressive estimate.

There are several plasma parameters such as its lifetime, skin-effects, and the inelastic collision frequency, which are relatively unknown and hard to model. To get a handle on these parameters first experimental efforts have been performed by measuring the radio scattering from the plasma induced in a block of ice which was irradiated by a beam of high-energy electrons. This experiment was performed at the Electron Light Source of the Telescope Array collaboration. Even-though first results indicate a possible scatter, a large amount of background from different sources was prominent in the signal region and no clear conclusion can be drawn. Nevertheless, we consider these initial results to be promising and conclude that they warrant further investigation.

\section{Acknowledgements}
The authors would like to thank the TA collaboration and the Chiba university Askaryan Radio Array group for their invaluable help obtaining the experimental results presented in these proceedings.

\end{document}